\documentstyle[amssymb,aps]{revtex}

\begin{document}
\draft
\title{Generalized Competing Glauber-type Dynamics and Kawasaki-type Dynamics}
\author{Han Zhu$^2$, Jian-yang Zhu$^{1,3\thanks{%
Author to whom correspondence should be addressed. Address correspondence to
Department of Physics, Beijing Normal University, Beijing 100875, China.
Email address: zhujy@bnu.edu.cn}}$ and Yang Zhou$^2$}
\address{$^1$CCAST (World Laboratory), Box 8730, Beijing 100080, China\\
$^2$Department of Physics, Nanjing University, Nanjing, 210093, China\\
$^3$Department of Physics, Beijing Normal University, Beijing 100875, China}
\date{\today }
\maketitle

\begin{abstract}
In this article, we have given a systematic formulation of the new
generalized competing mechanism: the Glauber-type single-spin transition
mechanism, with probability $p$, simulates the contact of the system with
the heat bath, and the Kawasaki-type spin-pair redistribution mechanism,
with probability $1-p$, simulates an external energy flux. These two
mechanisms are natural generalizations of Glauber's single-spin flipping
mechanism and Kawasaki's spin-pair exchange mechanism respectively. On the
one hand, the new mechanism is in principle applicable to arbitrary systems,
while on the other hand, our formulation is able to contain a mechanism that
just directly combines single-spin flipping and spin-pair exchange in their
original form.{\it \ }Compared with the conventional mechanism, the new
mechanism does not assume the simplified version and leads to greater
influence of temperature. The fact, order for lower temperature and disorder
for higher temperature, will be universally true. In order to exemplify this
difference, we applied the mechanism to $1D$ Ising model and obtained
analytical results. We also applied this mechanism to kinetic Gaussian model
and found that, above the critical point there will be only paramagnetic
phase, while below the critical point, the self-organization as a result of
the energy flux will lead the system to an interesting heterophase, instead
of the initially guessed antiferromagnetic phase. We studied this process in
details.
\end{abstract}


\section{Introduction}

In recent years, there have been continuing efforts towards a clear picture
of the self-organization phenomena in the phase transitions of magnetic
systems. Most of the works\cite{first,2,1dising,4,5,6,7,8} have been
concentrated on Ising and Ising-like systems, governed by two competing
dynamics: Glauber's single-spin flipping mechanism\cite{Glauber} and
Kawasaki's spin-pair exchange mechanism\cite{Kawasaki}, both with a
probability. The system is coupled to a heat bath at a given temperature
while being subject to an external energy flux. Glauber's flipping mechanism
is to simulate the contact of the system with the heat bath. Changing the
order parameter, it favors lower system energy. On the other hand,
Kawasaki's exchange mechanism is to simulate the influence of the energy
flux. Keeping the order parameter conserved, it favors higher system energy.
With these two competing mechanisms and the corresponding master equation,
people expect to obtain the evolution of the system. As exact treatment is
not possible for $2D$ Ising model, consequently Monte Carlo simulation and
methods as Dynamic Pair Approximation have been employed. The results
obtained helped to determine the interesting phase diagrams. (However,
people are surprised to find there contradictions in the predictions of MC
simulations and the dynamical mean-field theory, since they are both proved
successful to yield good qualitative results in other studies. Though
revisions of MC simulations are made and more accurate versions of
mean-field theory are presented, the puzzle still remains\cite{6}.)

In our earlier studies, we have presented single-spin transition mechanism%
\cite{sst1,sst2} and spin-pair redistribution mechanism\cite{ssr}. These two
dynamics are natural generalizations of Glauber's single-spin flipping
mechanism and Kawasaki's spin-pair exchange mechanism respectively. They
have similar mathematical expressions, and become counterparts of each other
in the non-conserved and conserved processes respectively. As an example of
the applications, we studied kinetic Gaussian model with both of them
respectively. Our study shows that, in translational-invariant lattices, 
{\it the dynamic critical exponent }$z=1/v=2${\it \ is independent of space
dimensionality and the governing dynamical mechanism. }Its dynamic
properties are summarized in Sec. \ref{Sec.4}.

In this article, we formulate the competing dynamics combining single-spin
transition and spin-pair redistribution. As these two mechanisms themselves
are universal, the combined one is also applicable to arbitrary systems, and
it can be deemed as a generalization. In Sec. \ref{Sec.2}, we first briefly
review the two mechanisms respectively and then give the formulation of the
competing mechanism. In Sec. \ref{Sec.3}, we explain the differences between
our mechanism and that adopted conventionally, taking $1D$ Ising model as an
example. In Sec. \ref{Sec.4}, as the chief task of this article, we apply it
to kinetic Gaussian model and report the findings. In Sec. \ref{Sec.5} we
summarize our study with some discussions.

\section{The competing mechanism}

\label{Sec.2}

First we briefly review the single-spin transition mechanism and spin-pair
redistribution mechanism.

\subsection{Single-spin transition mechanism}

Glauber's single-spin flipping mechanism allows an Ising system to evolve
with its spins flipping to their opposite. In single-spin transition
mechanism\cite{sst1,sst2}, a single spin $\sigma _i$ may change itself to
any possible values, $\hat{\sigma}_i$, and the master equation is 
\begin{equation}
\frac d{dt}P(\{\sigma \},t)=-\sum_i\sum_{\hat{\sigma}_i}\left[ W_i(\sigma
_i\rightarrow \hat{\sigma}_i)P(\{\sigma \},t)-W_i(\hat{\sigma}_i\rightarrow
\sigma _i)P(\{\sigma _{j\neq i}\},\hat{\sigma}_i,t)\right] ,
\label{sst-master}
\end{equation}
The transition probability is in a normalized form determined by a heat
Boltzmann factor, 
\begin{equation}
W_i(\sigma _i\rightarrow \hat{\sigma}_i)=\frac 1{Q_i}\exp \left[ -\beta 
{\cal H}_i\left( \hat{\sigma}_i,\sum_{<ij>}\sigma _j\right) \right]
,Q_i=\sum_{\hat{\sigma}_i}\exp \left[ -\beta {\cal H}_i\left( \hat{\sigma}%
_i,\sum_{<ij>}\sigma _j\right) \right] .  \label{sst-w}
\end{equation}
One can clearly see that this mechanism favors a lower Hamiltonian of the
system. Based on the master equation Eqs.(\ref{sst-master}), one can prove
that the time expectations of single-spin and correlation functions are 
\begin{eqnarray}
&&\frac d{dt}\left\langle \sigma _{i_1}(t)\sigma _{i2}(t)\cdots \sigma
_{i_n}(t)\right\rangle  \nonumber \\
&=&-n\left\langle \sigma _{i_1}\sigma _{i_2\cdots }\sigma
_{i_n}\right\rangle +\sum_{\{\sigma \}}\left\{ \sum_{k=1}^n\left[ \left(
\prod_{j(\neq k)=1}^n\sigma _{i_j}\right) \left( \sum_{\hat{\sigma}_{i_k}}%
\hat{\sigma}_{i_k}W_{i_k}(\sigma _{i_k}\rightarrow \hat{\sigma}%
_{i_k})\right) \right] \right\} P(\{\sigma \},t).  \label{Multispin C}
\end{eqnarray}
When $n=1$, it is 
\begin{equation}
\frac d{dt}q_k\left( t\right) =-q_k\left( t\right) +\sum_{\left\{ \sigma
\right\} }\left[ \sum_{\hat{\sigma}_k}\hat{\sigma}_kW_k\left( \sigma
_k\rightarrow \hat{\sigma}_k\right) \right] P\left( \left\{ \sigma \right\}
;t\right) .  \label{G-q}
\end{equation}

\subsection{Spin-pair redistribution mechanism}

Kawasaki's spin-pair exchange mechanism allows an Ising system to evolve
with its nearest neighbors exchanging their spin values. In spin-pair
redistribution mechanism\cite{ssr}, two neighboring spins, $\sigma _j\sigma
_l$, may change to any possible values, $\hat{\sigma}_j\hat{\sigma}_l$, as
long as their sum are conserved, and the master equation is 
\begin{eqnarray}
\frac d{dt}P(\{\sigma \},t) &=&\sum_{\left\langle jl\right\rangle }\sum_{%
\hat{\sigma}_j,\hat{\sigma}_l}\left[ -W_{jl}\left( \sigma _j\sigma
_l\rightarrow \hat{\sigma}_j\hat{\sigma}_l\right) P\left( \left\{ \sigma
\right\} ;t\right) \right.  \nonumber \\
&&\left. +W_{jl}\left( \hat{\sigma}_j\hat{\sigma}_l\rightarrow \sigma
_j\sigma _l\right) P\left( \left\{ \sigma _{i\neq j},\sigma _{l\neq
k}\right\} ,\hat{\sigma}_j,\hat{\sigma}_l;t\right) \right] .
\label{ssr-master}
\end{eqnarray}
The redistribution probability is also in a normalized form determined by a
heat Boltzmann factor, 
\begin{equation}
W_{jl}\left( \sigma _j\sigma _l\rightarrow \hat{\sigma}_j\hat{\sigma}%
_l\right) =\frac 1{Q_{jl}}\delta _{\sigma _j+\sigma _l,\hat{\sigma}_j+\hat{%
\sigma}_l}\exp \left[ -\beta {\cal H}_{jl}\left( \hat{\sigma}_j,\hat{\sigma}%
_l,\left\{ \sigma _m\right\} _{m\neq j,l}\right) \right] ,  \label{ssr-w}
\end{equation}
where the normalization factor $Q_{jl}$ is 
\[
Q_{jl}=\sum_{\hat{\sigma}_j,\hat{\sigma}_l}\delta _{\sigma _j+\sigma _l,\hat{%
\sigma}_j+\hat{\sigma}_l}\exp \left[ -\beta {\cal H}_{jl}\left( \hat{\sigma}%
_j,\hat{\sigma}_l,\left\{ \sigma _m\right\} _{m\neq j,l}\right) \right] .. 
\]
(Here it clearly favors a lower system Hamiltonian, but in the combined
mechanism we shall change the sign before $\beta {\cal H}_{jl}$ and make it
turn to the opposite.) The time expectation of single-spin is 
\begin{equation}
\frac d{dt}q_k\left( t\right) =-2dq_k\left( t\right) +\sum_w\sum_{\left\{
\sigma \right\} }\left[ \sum_{\hat{\sigma}_k,\hat{\sigma}_{k+w}}\hat{\sigma}%
_kW_{k,k+w}\left( \sigma _k\sigma _{k+w}\rightarrow \hat{\sigma}_k\hat{\sigma%
}_{k+w}\right) \right] P\left( \left\{ \sigma \right\} ;t\right) ,
\end{equation}
where $d$ is the space dimensionality and $\sum_w$ means a summation taken
over the nearest neighbors.

\subsection{The competing mechanism}

With the competing mechanisms, single-spin transition with probability $p$
and spin-pair redistribution with probability $1-p$, the master equation can
be written as, 
\begin{equation}
\frac d{dt}P\left( \left\{ \sigma \right\} ,t\right) =pG_{me}+\left(
1-p\right) K_{me},  \label{master eq}
\end{equation}
where the Glauber-type 
\begin{equation}
G_{me}=\sum_i\sum_{\hat{\sigma}_i}\left[ -W_i\left( \sigma _i\rightarrow 
\hat{\sigma}_i\right) P\left( \left\{ \sigma \right\} ,t\right) +W_i\left( 
\hat{\sigma}_i\rightarrow \sigma _i\right) P\left( \left\{ \sigma _{j\neq
i}\right\} ,\hat{\sigma}_i;t\right) \right] ,  \label{G-mq}
\end{equation}
and the Kawasaki-type 
\begin{eqnarray}
K_{me} &=&\sum_{\left\langle jl\right\rangle }\sum_{\hat{\sigma}_j,\hat{%
\sigma}_l}\left[ -W_{jl}\left( \sigma _j\sigma _l\rightarrow \hat{\sigma}_j%
\hat{\sigma}_l\right) P\left( \left\{ \sigma \right\} ;t\right) \right. 
\nonumber \\
&&\left. +W_{jl}\left( \hat{\sigma}_j\hat{\sigma}_l\rightarrow \sigma
_j\sigma _l\right) P\left( \left\{ \sigma _{i\neq j},\sigma _{l\neq
k}\right\} ,\hat{\sigma}_j,\hat{\sigma}_l;t\right) \right] .  \label{K-mq}
\end{eqnarray}
The Glauber-type mechanism is used to simulate the contact of the system
with the external heat bath, and the transition probability is of the form
given by Eqs.(\ref{sst-w}). This mechanism favors lower energy of the
system. The spin-pair redistribution mechanism is used to simulate the
energy flux. The redistribution probability given above favors a lower
system Hamiltonian, but what we need here is to the contrary. {\it We can
turn this tendency to its opposite if we change the sign before }$\beta 
{\cal H}_{jl}${\it \ in the redistribution probability.} It has the
following form, 
\begin{equation}
W_{jl}\left( \sigma _j\sigma _l\rightarrow \hat{\sigma}_j\hat{\sigma}%
_l\right) =\frac 1{Q_{jl}}\delta _{\sigma _j+\sigma _l,\hat{\sigma}_j+\hat{%
\sigma}_l}\exp \left[ \beta {\cal H}_{jl}\left( \hat{\sigma}_j,\hat{\sigma}%
_l,\left\{ \sigma _m\right\} _{m\neq j,l}\right) \right] ,  \label{red-pro}
\end{equation}
\[
Q_{jl}=\sum_{\hat{\sigma}_j,\hat{\sigma}_l}\delta _{\sigma _j+\sigma _l,\hat{%
\sigma}_j+\hat{\sigma}_l}\exp \left[ \beta {\cal H}_{jl}\left( \hat{\sigma}%
_j,\hat{\sigma}_l,\left\{ \sigma _m\right\} _{m\neq j,l}\right) \right] . 
\]
This normalized form implies that the tendency is toward a higher system
Hamiltonian. {\it The word, competition, is in fact that between the two
opposite directions either favored by one mechanism.}

As given above, we have already obtained the time expectation of single-spin
with either mechanism respectively. One can prove that with competing
mechanisms it will be, 
\begin{equation}
\frac d{dt}q_k\left( t\right) =pQ_k^G+\left( 1-p\right) Q_k^K,  \label{com-q}
\end{equation}
where the Glauber-type, 
\begin{equation}
Q_k^G=-q_k\left( t\right) +\sum_{\left\{ \sigma \right\} }\left[ \sum_{\hat{%
\sigma}_k}\hat{\sigma}_kW_k\left( \sigma _k\rightarrow \hat{\sigma}_k\right)
\right] P\left( \left\{ \sigma \right\} ;t\right) ,  \label{G-q-t}
\end{equation}
and the Kawasaki-type, 
\begin{equation}
Q_k^K=-2dq_k\left( t\right) +\sum_w\sum_{\left\{ \sigma \right\} }\left[
\sum_{\hat{\sigma}_k,\hat{\sigma}_{k+w}}\hat{\sigma}_kW_{k,k+w}\left( \sigma
_k\sigma _{k+w}\rightarrow \hat{\sigma}_k\hat{\sigma}_{k+w}\right) \right]
P\left( \left\{ \sigma \right\} ;t\right) .  \label{K-q-t}
\end{equation}
There are equations available for correlation functions with competing
mechanisms, but in later studies we find this single-spin equation is enough
to yield satisfying results.

\section{On Ising Model}

\label{Sec.3}

As mentioned above, these years great efforts have been contributed to Ising
model with competing dynamics. Because exact analytical treatment is too
hard, most of the studies have been either approximation or Monte Carlo
simulation. There are some differences between the conventional method and
ours. In the following paragraph we present our considerations.

The results one may expect directly depend on the expression of the
transition (flipping, exchange, redistribution) probabilities. We think that
there are two requirements: First, this probability should contain the
Hamiltonian, and thus naturally favors either higher energy (Kawasaki-type)
or lower energy (Glauber-type). Second, introducing temperature into it, we
require that the transition be influenced by the heat noise. It is the first
requirement that makes the two mechanisms compete and in all the studies it
has been well adopted. However, due to the difficulties of actual practice,
most of them used the simplified versions. In most of the studies, the
temperature factor has not been introduced into the exchange probability,
while in the flipping probability it has been only partly combined.
Typically for ferromagnetic Ising model it has been set as: Glauber-type 
\[
W_i=\min \left\{ 1,\exp \left( \vartriangle E_i/KT\right) \right\} , 
\]
Kawasaki-type 
\[
W_{ij}=\left\{ 
\begin{array}{c}
1,\vartriangle E_{ij}>0 \\ 
0,\vartriangle E_{ij}\leqslant 0
\end{array}
\right. . 
\]

In the new mechanism the transition and redistribution probabilities do not
take the simplified versions. Besides some mathematical aspects such as
normalization, their difference lies in the role of temperature. With the
new mechanism there is greater influence of heat on the system. The fact,
order for lower temperature and disorder for higher temperature, is not
universally true in the phase diagrams obtained in earlier studies\cite
{first,2,1dising,4,5,6,7}. However, we believe that this expectation will be
unshakable if the system is governed by the new mechanism. In order to
further study it, we applied our method to $1D$ Ising model. It is well
known that, due to the heat noise, there is only paramagnetic phase in $1D$
Ising model. The analytical results we obtained confirms this conclusion,
however one increases the energy flux, for all temperatures. (The details
are in Appendix A, but we suggest you read it later for your easier
understanding of our method.) $1D$ Ising model governed by the conventional
mechanism has been studied in Ref.\cite{1dising} with approximation method
and MC simulation. Further application on $2D$ Ising model is beyond the
scope of this article. There is no better or worse, since the two mechanisms
have different characteristics, but we think the comparison will be
interesting and also feasible in practice.

\section{On the kinetic Gaussian model}

\label{Sec.4}

In this section we apply the new mechanism to the $3D$ kinetic Gaussian
model and report our findings with the phase diagram. One and
two-dimensional cases are quite similar. First we briefly review some basic
properties of this model.

Gaussian model, proposed by T. H. Berlin and M. Kac, at first in order to
make Ising model more tractable, is an continuous-spin model. It has the
same Hamiltonian as the Ising model (three dimensional), 
\begin{equation}
-\beta {\cal H}=K\sum_{i,j,k=1}^N\sum_w\sigma _{ijk}\left( \sigma
_{i+w,jk}+\sigma _{ij+w,k}+\sigma _{ij,k+w}\right) ,  \label{G-Ham}
\end{equation}
where $K=J/k_BT$ and $\sum_w$ means summation over nearest neighbors.
Compared with the Ising model, it has two extensions: First, the spins $%
\sigma _{ijk}$ can take any real value between $\left( -\infty ,+\infty
\right) $. Second, to prevent the spins from tending to infinity, the
probability of finding a given spin between $\sigma _{ijk}$ and $\sigma
_{ijk}+d\sigma _{ijk}$ is assumed to be the Gaussian-type distribution 
\begin{equation}
f\left( \sigma _{ijk}\right) d\sigma _{ijk}=\sqrt{\frac b{2\pi }}\exp \left(
-\frac b2\sigma _{ijk}^2\right) d\sigma _{ijk},  \label{Spin-dis}
\end{equation}
where $b$ is a distribution constant independent of temperature. Being an
extension of Ising model, Gaussian model is quite different however. In the
equilibrium case, on translational invariant lattices it is exactly
solvable, and later as a starting point to study the unsolvable models it
has also been investigated with mean field theory and the momentum-space
renormalization-group method.

As an example of the applications of single-spin transition and spin-pair
redistribution mechanism, we have studied kinetic Gaussian model with both
of them respectively. We summarize its dynamic properties as follows\cite
{sst1,sst2,ssr}. The inherent dynamical competition of this model is that:
the system tries to lower its Hamiltonian with the spins tending to
infinity, while the Gaussian-type probability serves to restrict this
tendency. Above the critical temperature, the prevailing heat noise permits
only a disordered state, whereas below the critical point some kind of order
will appear. Our study shows that, on translational-invariant lattices, {\it %
the dynamic critical exponent }$z=1/v=2${\it \ is independent of space
dimensionality and the governing dynamical mechanism.}

Now we turn to treat the $3D$ kinetic Gaussian model with the competing
dynamics. $1D$ and $2D$ systems can be treated in a similar way and they
have qualitatively the same properties. In earlier studies we obtained the
time expectation of single-spin. With the competing mechanism we can borrow
these equations from Ref.\cite{sst1} and \cite{ssr}.

(1) With Glauber dynamics: 
\begin{equation}
\frac d{dt}q_{ijk}\left( t\right) \equiv Q_{ijk}^G=-q_{ijk}+\frac Kb\left(
q_{i-1,j,k}+q_{i+1,j,k}+q_{i,j-1,k}+q_{i,j+1,k}+q_{i,j,k-1}+q_{i,j,k+1}%
\right) .
\end{equation}

(2) With Kawasaki dynamics: since we have changed the sign before system
Hamiltonian in the redistribution probability, 
\[
-\beta {\cal H}=K\sum_{i,j,k=1}^N\sum_w\sigma _{ijk}\left( \sigma
_{i+w,jk}+\sigma _{ij+w,k}+\sigma _{ij,k+w}\right) , 
\]
\[
\Longrightarrow \beta {\cal H}=\left( -K\right) \sum_{i,j,k=1}^N\sum_w\sigma
_{ijk}\left( \sigma _{i+w,jk}+\sigma _{ij+w,k}+\sigma _{ij,k+w}\right) , 
\]
all the expressions will remain the same if we switch $K$ to $-K$. Here we
will have to do the same, 
\begin{eqnarray}
\frac d{dt}q_{ijk}\left( t\right) &\equiv &Q_{ijk}^K=\frac 1{2\left[
b+\left( -K\right) \right] }b\left\{ \left[ \left(
q_{i+1,j,k}-q_{ijk}\right) -\left( q_{ijk}-q_{i-1,j,k}\right) \right] \right.
\nonumber \\
&&\left. +\left[ \left( q_{i,j+1,k}-q_{ijk}\right) -\left(
q_{ijk}-q_{i,j-1,k}\right) \right] +\left[ \left( q_{i,j,k+1}-q_{ijk}\right)
-\left( q_{ijk}-q_{i,j,k-1}\right) \right] \right\}  \nonumber \\
&&+\frac{\left( -K\right) }{2\left[ b+\left( -K\right) \right] }\left[
2\left( 2q_{i-1,j,k}-q_{i-1,j+1,k}-q_{i-1,j-1,k}\right) +\left(
2q_{i-1,j,k}-q_{ijk}-q_{i-2,j,k}\right) \right.  \nonumber \\
&&+2\left( 2q_{i+1,j,k}-q_{i+1,j+1,k}-q_{i+1,j-1,k}\right) +\left(
2q_{i+1,j,k}-q_{ijk}-q_{i+2,j,k}\right)  \nonumber \\
&&+2\left( 2q_{i,j-1,k}-q_{i,j-1,k+1}-q_{i,j-1,k-1}\right) +\left(
2q_{i,j-1,k}-q_{ijk}-q_{i,j-2,k}\right)  \nonumber \\
&&+2\left( 2q_{i,j+1,k}-q_{i,j+1,k+1}-q_{i,j+1,k-1}\right) +\left(
2q_{i,j+1,k}-q_{ijk}-q_{i,j+2,k}\right)  \nonumber \\
&&+2\left( 2q_{i,j,k-1}-q_{i-1,j,k-1}-q_{i+1,j,k-1}\right) +\left(
2q_{i,j,k-1}-q_{ijk}-q_{i,j,k-2}\right)  \nonumber \\
&&\left. +2\left( 2q_{i,j,k+1}-q_{i+1,j,k+1}-q_{i-1,j,k+1}\right) +\left(
2q_{i,j,k+1}-q_{ijk}-q_{i,j,k+2}\right) \right] .
\end{eqnarray}
This expression is in fact not as complex as it seems. Each term in a
bracket is a second order derivative of $q$, and they will cancel each other
if a summation is taken over all the spins. Thus $\sum_{ijk}Q_{ijk}^K=0$.

Then with the competing mechanisms, 
\begin{equation}
\frac d{dt}q_{ijk}\left( t\right) =pQ_{ijk}^G+\left( 1-p\right) Q_{ijk}^K.
\label{1}
\end{equation}

It is natural to first consider using the system Hamiltonian, 
\[
{\cal H}=-J\sum_{\left\langle i,j\right\rangle }\sigma _i\sigma _j, 
\]
to characterize its behavior. If we find the Hamiltonian is decreasing, then
the system is believed to be evolving to Ferromagnetic phase; if the
Hamiltonian is stable around zero, the system must be in Paramagnetic phase;
if the Hamiltonian is increasing, then the system is evolving towards
Antiferromagnetic phase. Unfortunately, in Gaussian model although we can
conveniently obtain the exact result for $\left\langle \sigma _i\sigma
_j\right\rangle $ and the average value, $\sum_{ij}\left\langle \sigma
_i\sigma _j\right\rangle /N^2$, we find it difficult to obtain an analytical
result of the system Hamiltonian. In order to differentiate between these
phases, instead we study the following two aspects. First, magnetization, 
\begin{equation}
M\left( t\right) =\frac 1N\sum_kq_k\left( t\right) ,  \label{2}
\end{equation}
and second, 
\begin{equation}
M^{\prime }\left( t\right) =\frac 1N\sum_{ijk}q_{ijk}^{\prime }\left(
t\right) =\frac 1N\sum_{ijk}\left( -\right) ^{i+j+k}q_{ijk}\left( t\right) .
\label{3}
\end{equation}
For pure ferromagnetic phase, if without any extra conditions, we expect
nonzero magnetization, that is to say, $M\left( t\right) \neq 0$. At the
same time, if the system is divided into two penetrating subsets, in one of
which the spins have $i+j+k$ being an odd number, and in the other one the
spins have $i+j+k$ being even, then one will find these two subsets are
almost identical, and that leads to $M^{\prime }\left( t\right) =0$. So $%
M\left( t\right) \neq 0$ and $M^{\prime }\left( t\right) =0$ is the
characteristic of ferromagnetic phase. For pure antiferromagnetic phase, we
expect such a situation that the system can be divided into two penetrating
opposing sublattices in the way mentioned above, one of positive spin and
one of negative. Except the direction of the spins, these two sublattices
are identical. This leads to $M^{\prime }\left( t\right) \neq 0$ but $%
M\left( t\right) =0$. In disordered paramagnetic phase, we expect disorder
of the whole system and both $M\left( t\right) $ and $M^{\prime }\left(
t\right) $ to be zero. Thus, if we get the evolution of $M\left( t\right) $
and $M^{\prime }\left( t\right) $, we can decide in which phase the system
is.

With Eqs.(\ref{1}), (\ref{2}) and (\ref{3}) we can write the time derivative
of $M\left( t\right) $ and $M^{\prime }\left( t\right) $ as, 
\begin{eqnarray}
\frac{dM\left( t\right) }{dt} &=&\frac 1N\sum_k\frac d{dt}q_k\left( t\right)
\nonumber \\
&=&p\left( -1+\frac{6K}b\right) M\left( t\right) .  \label{M}
\end{eqnarray}
and 
\begin{eqnarray}
\frac{dM^{\prime }\left( t\right) }{dt} &=&\frac 1N\sum_k\frac d{dt}%
q_k^{\prime }\left( t\right)  \nonumber \\
&=&\left[ -p\left( 1+\frac{6K}b\right) +6\left( 1-p\right) \frac{6K-b}{b-K}%
\right] M^{\prime }\left( t\right)  \label{M'}
\end{eqnarray}
The details of Eqs.(\ref{M'}) are in Appendix B. Because we want $K/b$ to be
positive, the inequality 
\[
-p\left( 1+\frac{6K}b\right) +6\left( 1-p\right) \frac{6K-b}{b-K}>0, 
\]
becomes 
\begin{equation}
\frac 1{12p}\left( -36+41p+\sqrt{\left( 1296-2808p+1561p^2\right) }\right) <%
\frac Kb<1,
\end{equation}
for $0<p<1$.

Fig.1 shows the phase diagram of $3D$ kinetic Gaussian model. It is divided
into several regions by the following three curves, 
\begin{equation}
\exp \left( -K/b\right) =e^{-1/6},  \label{curve1}
\end{equation}
\begin{equation}
\exp \left( -K/b\right) =1/e,  \label{curve2}
\end{equation}
and 
\begin{equation}
\exp \left( -K/b\right) =\exp \left[ -\frac 1{12p}\left( -36+41p+\sqrt{%
\left( 1296-2808p+1561p^2\right) }\right) \right] .  \label{curve3}
\end{equation}
Phase Diagrams for $1D$ and $2D$ models are quite similar.

We know that the critical point of $3D$ Gaussian model is $K_c=b/2d$, where $%
d$ is the space dimensionality. In the region above line (\ref{curve1}), the
temperature is higher than the critical value. Both $M\left( t\right) $ and $%
M^{\prime }\left( t\right) $ are approaching zero exponentially, and this is
identified as paramagnetic phase. The overwhelming heat noise permits no
observable magnetization, and the Hamiltonian is static in equilibrium. When
the temperature is below $T_c$, the heat noise becomes secondary and some
kind of order appears.

For the region below line (\ref{curve1}) and above curve (\ref{curve3}), as
well as that below line (\ref{curve2}), we have exponentially increasing $%
M\left( t\right) $ while $M^{\prime }\left( t\right) $ is approaching zero,
and this corresponds to ferromagnetic phase. In this phase, Glauber-type
mechanism prevails and the energy is decreasing. There is observable
homogeneous magnetization, the direction of which depends on that of the
initial magnetization.

In the region between curve (\ref{curve2}) and line (\ref{curve3}) we have
both exponentially increasing $M\left( t\right) $ and $M^{\prime }\left(
t\right) $, and it can not be simply identified as ferromagnetic or
antiferromagnetic. We give it a name, heterophase. In this region, the
Kawasaki mechanism and the energy flux control the system, and the energy
has a tendency to increase. The spin values are interesting. Our analytical
results show that $M\left( t\right) =M\left( 0\right) e^{At}$ and $M^{\prime
}\left( t\right) =M^{\prime }\left( 0\right) e^{Bt}$. This means that, if
the initial phase is ferromagnetic, $M^{\prime }\left( 0\right) =0$, but $%
M\left( 0\right) \neq 0$, later we will observe $\left| M\left( t\right)
\right| $ increasing but $M^{\prime }\left( t\right) $ staying at zero. If
initially the system is antiferromagnetic, $M\left( 0\right) =0$, but $%
M^{\prime }\left( 0\right) \neq 0$, later we will have $\left| M^{\prime
}\left( t\right) \right| $ increasing but $M\left( t\right) $ staying at
zero. If initially a disordered paramagnetic phase is given, $M\left(
0\right) =0$, and also $M^{\prime }\left( 0\right) =0$, we will get both
zero $M\left( t\right) $ and $M^{\prime }\left( t\right) $. From this one
may get confused,---what is the real picture? How is the energy sure to
increase if the spin values are dependent on other conditions? Direct
computer simulation reveals the key. For example, we apply the periodic
boundary condition and initially set the system as, 
\begin{eqnarray*}
&&\ldots +1,+1,-1,+1,+1,-1,+1,+1,-1,+1, \\
&&+1,-1,+1,+1,-1,+1,+1,-1,+1,+1,\ldots
\end{eqnarray*}
This leads to $M\left( 0\right) \neq 0$, but $M^{\prime }\left( 0\right) =0$%
. After $1$ second, it becomes (we only give approximate values), 
\begin{eqnarray*}
&&\ldots 0.7,1.1,0.2,0.8,0.6,0.3,0.6,0.6,0.2,0.6, \\
&&0.6,0.2,0.6,0.6,0.3,0.6,0.8,0.2,1.1,0.7,\ldots
\end{eqnarray*}
After $10$ seconds, 
\begin{eqnarray*}
&&\ldots 9,13,5,17,3,17,3,15,6,11, \\
&&11,6,15,3,17,3,17,5,13,9,\ldots
\end{eqnarray*}
After $24$ seconds, 
\begin{eqnarray*}
&&\ldots -249,6941,-6735,12089,-10041,13228,-8902,9922,-3755,3435, \\
&&3435,-3755,9922,-8902,13228,-10041,12089,-6735,6941,-249,\ldots
\end{eqnarray*}
This shows the routine of {\it self-organization} the system chooses in this
specific case, and it is a self-explanatory picture of how the energy
manages to increase while $M^{\prime }\left( t\right) =0$.

However, there is one exception. If at the beginning the system is set in a
pure homogeneous ferromagnetic phase, we will find that the system remains
in this homogeneous state. How can one explain this? Actually the system
does have a tendency to increase its energy in this region, but it needs a
hint to know exactly how. Here we cite the words of Tome and Oliveira\cite
{first}: It is ''the result, in the case considered here, of a far from
equilibrium process, namely, the continuous flux of energy into the system.
Thus an instability of the usual (equilibrium) solutions leads the system
toward states with spatial self-organized structure.'' The self-organization
routine is closely related with a basic phenomenon of symmetry loss. A
system under conditions that lead to paramagnetic phase is like a ball
placed on the ground, for which all directions are identical. A system set
in such a heterophase is like a ball placed on the top of another sphere.
The upper ball has a tendency to fall down, but the direction depends on
some kind of disturbance. If there is no disturbance at all, it will stay
still. It will even have its energy increasing if the lower ball is being
lifted up. Surely in reality we have never seen a ball stay still on the top
of another, because disturbance can not be avoided. So discussion here is
only to help qualitatively explain this strange phase and the mechanism
hidden behind.

\section{Results and some Discussion}

\label{Sec.5}

In this article, we have given a systematic formulation of the new competing
mechanism. The Glauber-type single-spin transition mechanism with
probability $p$ simulates the contact of the system with the heat bath and
the Kawasaki-type spin-pair redistribution mechanism with probability $1-p$
simulates an external energy flux. These two mechanisms themselves are
natural generalizations of Glauber's single-spin flipping mechanism and
Kawasaki's spin-pair exchange mechanism. Thus, on the one hand, this
mechanism is in principle applicable to arbitrary systems, while on the
other hand, our formulation is able to contain a mechanism that just {\it %
directly} combines single-spin flipping and spin-pair exchange {\it in their
original forms }(not simplified).{\it \ }Compared with the conventional one,
the new mechanism does not assume the simplified versions. Their difference
lies in the different role the system temperature plays. (As we have
emphasized before, there is no better or worse, since they are different.)
We applied the new mechanism to $1D$ Ising model and used the analytical
results to exemplify this difference. With the new mechanism, there is
greater influence of temperature and the fact, order for lower temperature
and disorder for higher temperature, will be universally true.

In Sec. \ref{Sec.4}, we applied this mechanism to $3D$ kinetic Gaussian
model. $1D$ and $2D$ models can be treated following a similar way and have
qualitatively the same properties. We used $M\left( t\right) $ and $%
M^{\prime }\left( t\right) $ (their definition given in that section) to
characterize the system. In the phase diagram of $3D$ model, we confirm the
expectation, order for lower temperature and disorder for higher
temperature. For temperature above $T_c$, the system evolves to paramagnetic
phase. For temperature lower than $T_c$, we observe ferromagnetic phase and
another heterophase (instead of the antiferromagnetic phase as guessed).
This interesting heterophase is the result of the energy flux and
self-organization. We have analyzed it in details in that section, and hope
it will help make clear the self-organization process in phase transitions.
With regard to Ising model, we also hope it will be a good reference when
people are trying to solve the puzzling differences between the results
yielded by approximation method and MC simulation.

\acknowledgments 

This work was supported by the National Natural Science Foundation of China
under Grant No. 10075025.

\appendix 

\section{Calculational details on the kinetic Ising model}

We apply the new mechanism to $1D$ Ising model. We derive the time
expectation of single-spin using Eqs.(\ref{com-q}), 
\[
\frac d{dt}q_k\left( t\right) =pQ_k^G+\left( 1-p\right) Q_k^K. 
\]
In this combined form, the Glauber-type term is as Eqs.(\ref{G-q-t}), 
\begin{equation}
Q_k^G=-q_k\left( t\right) +\sum_{\left\{ \sigma \right\} }\left[ \sum_{\hat{%
\sigma}_k}\hat{\sigma}_kW_k\left( \sigma _k\rightarrow \hat{\sigma}_k\right)
\right] P\left( \left\{ \sigma \right\} ;t\right) ,  \label{App-1}
\end{equation}
and the Kawasaki-type is as Eqs.(\ref{K-q-t}), 
\begin{eqnarray}
Q_k^K &=&-2q_k\left( t\right) +\sum_{\left\{ \sigma \right\} }\left[ \sum_{%
\hat{\sigma}_k,\hat{\sigma}_{k+1}}\hat{\sigma}_kW_{k,k+1}\left( \sigma
_k\sigma _{k+1}\rightarrow \hat{\sigma}_k\hat{\sigma}_{k+1}\right) \right.
\label{App-2} \\
&&\left. +\sum_{\hat{\sigma}_k,\hat{\sigma}_{k-1}}\hat{\sigma}%
_kW_{k,k-1}\left( \sigma _k\sigma _{k-1}\rightarrow \hat{\sigma}_k\hat{\sigma%
}_{k-1}\right) \right] P\left( \left\{ \sigma \right\} ;t\right) .
\end{eqnarray}
First we calculate (\ref{App-1}). 
\begin{eqnarray}
&&\sum_{\hat{\sigma}_k}\hat{\sigma}_kW_k\left( \sigma _k\rightarrow \hat{%
\sigma}_k\right)  \nonumber \\
&=&\frac{\sigma _k\exp \left[ K\sigma _k\left( \sigma _{k-1}+\sigma
_{k+1}\right) \right] -\sigma _k\exp \left[ -K\sigma _k\left( \sigma
_{k-1}+\sigma _{k+1}\right) \right] }{\exp \left[ K\sigma _k\left( \sigma
_{k-1}+\sigma _{k+1}\right) \right] +\exp \left[ -K\sigma _k\left( \sigma
_{k-1}+\sigma _{k+1}\right) \right] }  \nonumber \\
&=&\left\{ 
\begin{array}{c}
0,\sigma _{k-1}=-\sigma _{k+1} \\ 
\tanh 2K,\sigma _{k-1}=\sigma _{k+1}=1 \\ 
-\tanh 2K,\sigma _{k-1}=\sigma _{k+1}=-1
\end{array}
\right.  \nonumber \\
&=&\frac 12\left( \sigma _{k-1}+\sigma _{k+1}\right) \tanh 2K.  \label{App-3}
\end{eqnarray}
So 
\begin{eqnarray}
Q_G^k &=&-q_k\left( t\right) +\sum_{\left\{ \sigma \right\} }\left[ \sum_{%
\hat{\sigma}_k}\hat{\sigma}_kW_k\left( \sigma _k\rightarrow \hat{\sigma}%
_k\right) \right] P\left( \left\{ \sigma \right\} ;t\right)  \nonumber \\
&=&-q_k\left( t\right) +\frac 12\left( q_{k-1}+q_{k+1}\right) \tanh 2K.
\label{App-4}
\end{eqnarray}
Second, we calculate (\ref{App-2}), 
\begin{eqnarray*}
&&\sum_{\hat{\sigma}_k,\hat{\sigma}_{k+1}}\hat{\sigma}_kW_{k,k+1}\left(
\sigma _k\sigma _{k+1}\rightarrow \hat{\sigma}_k\hat{\sigma}_{k+1}\right) \\
&=&\frac{\sigma _k\exp \left[ -K\left( \sigma _{k-1}\sigma _k+\sigma
_{k+1}\sigma _{k+2}\right) \right] +\sigma _{k+1}\exp \left[ -K\left( \sigma
_{k-1}\sigma _{k+1}+\sigma _k\sigma _{k+2}\right) \right] }{\exp \left[
-K\left( \sigma _{k-1}\sigma _k+\sigma _{k+1}\sigma _{k+2}\right) \right]
+\exp \left[ -K\left( \sigma _{k-1}\sigma _{k+1}+\sigma _k\sigma
_{k+2}\right) \right] } \\
&=&\left\{ 
\begin{array}{c}
\left( \sigma _k+\sigma _{k+1}\right) /2,\sigma _k=\sigma _{k+1} \\ 
\tanh \left[ -K\left( \sigma _{k-1}-\sigma _{k+2}\right) \right] ,\sigma
_k=-\sigma _{k+1}
\end{array}
\right. \\
&=&\frac{\sigma _k+\sigma _{k+1}}2+\frac{\left( \sigma _k-\sigma
_{k+1}\right) ^2}4 \\
&&\times \left[ \frac{\left( \sigma _{k-1}-1\right) \left( \sigma
_{k+2}+1\right) }4+\frac{\left( \sigma _{k-1}+1\right) \left( \sigma
_{k+2}-1\right) }{-4}\right] \tanh \left( -2K\right) . \\
&=&\frac{\sigma _k+\sigma _{k+1}}2+\frac 14\left( \sigma _{k-1}-\sigma
_{k+2}-\sigma _{k-1}\sigma _k\sigma _{k+1}+\sigma _k\sigma _{k+1}\sigma
_{k+2}\right) \tanh \left( -2K\right)
\end{eqnarray*}
And similarly, 
\begin{eqnarray*}
&&\sum_{\hat{\sigma}_k,\hat{\sigma}_{k-1}}\hat{\sigma}_kW_{k,k-1}\left(
\sigma _k\sigma _{k-1}\rightarrow \hat{\sigma}_k\hat{\sigma}_{k-1}\right) \\
&=&\frac{\sigma _k+\sigma _{k-1}}2+\frac 14\left( \sigma _{k+1}-\sigma
_{k-2}-\sigma _{k+1}\sigma _k\sigma _{k-1}+\sigma _k\sigma _{k-1}\sigma
_{k-2}\right) \tanh \left( -2K\right) .
\end{eqnarray*}
So 
\begin{eqnarray}
Q_k^K &=&-2q_k\left( t\right) +\sum_{\left\{ \sigma \right\} }\left[ \sum_{%
\hat{\sigma}_k,\hat{\sigma}_{k+1}}\hat{\sigma}_kW_{k,k+1}\left( \sigma
_k\sigma _{k+1}\rightarrow \hat{\sigma}_k\hat{\sigma}_{k+1}\right) \right. 
\nonumber \\
&&\left. +\sum_{\hat{\sigma}_k,\hat{\sigma}_{k-1}}\hat{\sigma}%
_kW_{k,k-1}\left( \sigma _k\sigma _{k-1}\rightarrow \hat{\sigma}_k\hat{\sigma%
}_{k-1}\right) \right] P\left( \left\{ \sigma \right\} ;t\right)  \nonumber
\\
&=&\frac 12\left( q_{k-1}+q_{k+1}-2q_k\right) +\frac 14\tanh \left(
-2K\right) \left( q_{k+1}-q_{k-2}+q_{k-1}-q_{k+2}\right)  \nonumber \\
&&+\frac 14\tanh \left( -2K\right) \left( \left\langle \sigma _k\sigma
_{k+1}\sigma _{k+2}\right\rangle -2\left\langle \sigma _{k+1}\sigma _k\sigma
_{k-1}\right\rangle +\left\langle \sigma _k\sigma _{k-1}\sigma
_{k-2}\right\rangle \right) .  \label{App-5}
\end{eqnarray}
Thus, combining (\ref{App-4}) and (\ref{App-5}) one will get, 
\begin{eqnarray}
\frac d{dt}q_k\left( t\right) &=&p\left[ -q_k\left( t\right) +\frac 12\left(
q_{k-1}+q_{k+1}\right) \tanh 2K\right]  \nonumber \\
&&+\left( 1-p\right) \left[ \frac 12\left( q_{k-1}+q_{k+1}-2q_k\right) +%
\frac 14\tanh \left( -2K\right) \left(
q_{k+1}-q_{k-2}+q_{k-1}-q_{k+2}\right) \right.  \nonumber \\
&&\left. +\frac 14\tanh \left( -2K\right) \left( \left\langle \sigma
_k\sigma _{k+1}\sigma _{k+2}\right\rangle -2\left\langle \sigma _{k+1}\sigma
_k\sigma _{k-1}\right\rangle +\left\langle \sigma _k\sigma _{k-1}\sigma
_{k-2}\right\rangle \right) \right] .  \label{1disingq}
\end{eqnarray}
In order to decide in which phase the system is, we have suggested two
quantities in Sec. \ref{Sec.4}, 
\[
M\left( t\right) =\sum_kq_k\left( t\right) , 
\]
and 
\[
M^{\prime }\left( t\right) =\sum_k\left( -\right) ^kq_k\left( t\right)
\equiv \sum_kq_k^{\prime }\left( t\right) . 
\]
We use their evolving tendency to characterize the system behavior.
Obviously, (1) in a homogeneous ferromagnetic phase, we expect $\left|
M\left( t\right) \right| /N\rightarrow 1$ and $\left| M^{\prime }\left(
t\right) \right| /N\rightarrow 0$; (2) in an antiferromagnetic phase
consisting of two penetrating and opposing sublattices, we will have $\left|
M^{\prime }\left( t\right) \right| /N\rightarrow 1$ and $\left| M\left(
t\right) \right| /N\rightarrow 0$; (3) in disordered paramagnetic phase,
both $M^{\prime }\left( t\right) /N$ and $M\left( t\right) /N$ will approach
zero. (A detailed analysis can be found in that section.) With Eqs.(\ref
{1disingq}) we can get, 
\[
\frac d{dt}M\left( t\right) \equiv \sum_k\frac d{dt}q_k\left( t\right)
=-p\left( 1-\tanh 2K\right) M\left( t\right) . 
\]
Thus 
\begin{equation}
M\left( t\right) =M\left( 0\right) \exp \left[ -p\left( 1-\tanh 2K\right)
t\right] .  \label{isingm}
\end{equation}
At the same time, 
\begin{eqnarray*}
&&\frac d{dt}\left( -\right) ^kq_k\left( t\right) \\
&\equiv &\frac d{dt}q_k^{\prime }\left( t\right) =p\left[ -q_k^{\prime
}\left( t\right) +\frac 12\left( -q_{k-1}^{\prime }-q_{k+1}^{\prime }\right)
\tanh 2K\right] \\
&&+\left( 1-p\right) \left[ \frac 12\left( -q_{k-1}^{\prime
}-q_{k+1}^{\prime }-2q_k^{\prime }\right) +\frac 14\tanh \left( -2K\right)
\left( -q_{k+1}^{\prime }-q_{k-2}^{\prime }-q_{k-1}^{\prime
}-q_{k+2}^{\prime }\right) \right. \\
&&\left. +\frac 14\tanh \left( -2K\right) \left( -\left\langle \sigma
_k^{\prime }\sigma _{k+1}^{\prime }\sigma _{k+2}^{\prime }\right\rangle
+2\left\langle \sigma _{k+1}^{\prime }\sigma _k^{\prime }\sigma
_{k-1}^{\prime }\right\rangle -\left\langle \sigma _k^{\prime }\sigma
_{k-1}^{\prime }\sigma _{k-2}^{\prime }\right\rangle \right) \right] ,
\end{eqnarray*}
and 
\begin{eqnarray*}
M^{\prime }\left( t\right) &\equiv &\sum_k\frac d{dt}q_k^{\prime }\left(
t\right) =\left[ p\left( -1-\tanh 2K\right) +\left( 1-p\right) \left(
-2+\tanh 2K\right) \right] M^{\prime }\left( t\right) \\
&=&\left[ -2+p+\left( 1-2p\right) \tanh 2K\right] M^{\prime }\left( t\right)
.
\end{eqnarray*}
Thus 
\begin{equation}
M^{\prime }\left( t\right) =M^{\prime }\left( 0\right) \exp \left\{ \left[
-2+p+\left( 1-2p\right) \tanh 2K\right] t\right\} .  \label{isingm'}
\end{equation}

In $1D$ Ising model, as given in Eqs.(\ref{isingm}) and (\ref{isingm'}), $%
M\left( t\right) $ and $M^{\prime }\left( t\right) $ are both approaching
zero exponentially, and we can make the conclusion that, however one
increases the energy flux, the system will stay in paramagnetic phase at
arbitrary temperature.

\section{Calculational details of Eqs.(\ref{M'})}

Kawasaki-type: 
\begin{eqnarray}
\frac d{dt}q_{ijk}^{\prime }\left( t\right) &=&\frac 1{2\left( b-K\right) }%
b\left\{ \left[ \left( -q_{i,j,k+1}^{\prime }-q_{ijk}^{\prime }\right)
-\left( q_{ijk}^{\prime }+q_{i,j,k-1}^{\prime }\right) \right] \right. 
\nonumber \\
&&\left. +\left[ \left( -q_{i+1,j,k}^{\prime }-q_{ijk}^{\prime }\right)
-\left( q_{ijk}^{\prime }+q_{i-1,j,k}^{\prime }\right) \right] +\left[
\left( -q_{i,j+1,k}^{\prime }-q_{ijk}^{\prime }\right) -\left(
q_{ijk}^{\prime }+q_{i,j-1,k}^{\prime }\right) \right] \right\}  \nonumber \\
&&+\frac{-K}{2\left( b-K\right) }\left[ 2\left( -2q_{i-1,j,k}^{\prime
}-q_{i-1,j+1,k}^{\prime }-q_{i-1,j-1,k}^{\prime }\right) +\left(
-2q_{i-1,j,k}^{\prime }-q_{ijk}^{\prime }-q_{i-2,j,k}^{\prime }\right)
\right.  \nonumber \\
&&+2\left( -2q_{i+1,j,k}^{\prime }-q_{i+1,j+1,k}^{\prime
}-q_{i+1,j-1,k}^{\prime }\right) +\left( -2q_{i+1,j,k}^{\prime
}-q_{ijk}^{\prime }-q_{i+2,j,k}^{\prime }\right)  \nonumber \\
&&+2\left( -2q_{i,j-1,k}^{\prime }-q_{i,j-1,k+1}^{\prime
}-q_{i,j-1,k-1}^{\prime }\right) +\left( -2q_{i,j-1,k}^{\prime
}-q_{ijk}^{\prime }-q_{i,j-2,k}^{\prime }\right)  \nonumber \\
&&+2\left( -2q_{i,j+1,k}^{\prime }-q_{i,j+1,k+1}^{\prime
}-q_{i,j+1,k-1}^{\prime }\right) +\left( -2q_{i,j+1,k}^{\prime
}-q_{ijk}^{\prime }-q_{i,j+2,k}^{\prime }\right)  \nonumber \\
&&+2\left( -2q_{i,j,k-1}^{\prime }-q_{i-1,j,k-1}^{\prime
}-q_{i+1,j,k+1}^{\prime }\right) +\left( -2q_{i,j,k-1}^{\prime
}-q_{ijk}^{\prime }-q_{i,j,k-2}^{\prime }\right)  \nonumber \\
&&\left. +2\left( -2q_{i,j,k+1}^{\prime }-q_{i+1,j,k+1}^{\prime
}-q_{i-1,j,k+1}^{\prime }\right) +\left( -2q_{i,j,k+1}^{\prime
}-q_{ijk}^{\prime }-q_{i,j,k+2}^{\prime }\right) \right] .
\end{eqnarray}
Glauber-type: 
\begin{equation}
\frac d{dt}q_{ijk}^{\prime }\left( t\right) =-q_{ijk}^{\prime }+\frac Kb%
\left( -q_{i-1,j,k}^{\prime }-q_{i+1,j,k}^{\prime }-q_{i,j+1,k}^{\prime
}-q_{i,j-1,k}^{\prime }-q_{i,j,k+1}^{\prime }-q_{i,j,k+1}^{\prime }\right) .
\end{equation}
And then 
\begin{eqnarray*}
\frac{dM^{\prime }\left( t\right) }{dt} &=&\frac 1N\sum_k\frac d{dt}%
q_k^{\prime }\left( t\right) \\
&=&\left[ -p\left( 1+\frac{6K}b\right) +\left( 1-p\right) 6\frac{6K-b}{b-k}%
\right] M^{\prime }\left( t\right) .
\end{eqnarray*}

\null\vskip0.2cm

\centerline{\bf Caption of figures} \vskip1cm

Fig.1. The phase diagram of $3D$ Gaussian model with competing dynamics,
Glauber-type with probability $p$, and Kawasaki-type with probability $1-p$.
The system exhibits paramagnetic phase (Para), Ferromagnetic phase (Ferro),
and heterophase (Hetero).

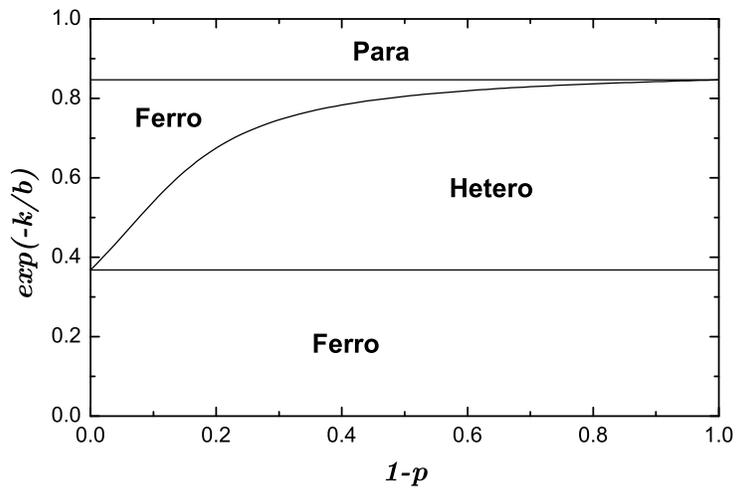
\end{document}